\documentclass{medphyspaper}

\usepackage{amsmath}
\usepackage{blindtext}
\usepackage{graphicx}
\usepackage{booktabs}
\usepackage{hyperref}
\usepackage{pgfplots} 
\usepackage{pgfgantt}
\usepackage{pdflscape}
\usepackage{romannum}
\usepackage{siunitx}
\usepackage{subcaption}
\usepackage{tikz}
\usepackage[utf8]{inputenc}
\usepackage{url}
\usepackage{todonotes}

\usepgfplotslibrary{patchplots}
\pgfplotsset{compat=newest} 
\pgfplotsset{plot coordinates/math parser=false}
\pgfplotsset{every axis/.append style={
		legend style={font=\scriptsize},
		title style={font=\scriptsize},
		label style={font=\scriptsize},
		tick label style={font=\scriptsize}
}}

\usetikzlibrary{external}

\DeclareMathOperator*{\argmax}{argmax}

\newlength{\fwidth}
\DeclareSIUnit{\MeVperU}{\mega\electronvolt\per\atomicmassunit}
\DeclareSIUnit\atomicmassunit{u}

\definecolor{myblue}{HTML}{0072BD} 
\definecolor{myorange}{HTML}{D95319} 
\definecolor{myyelllow}{HTML}{EDB120} 
\definecolor{mypurple}{HTML}{7E2F8E} 
\definecolor{mygreen}{HTML}{77AC30} 
\definecolor{mylightblue}{HTML}{4DBEEE} 
\definecolor{myred}{HTML}{A2142F} 

\title{Helium Range Viability for Online Range Probing in Mixed Carbon-Helium Beams}
\runningtitle{Helium Range in Mixed Beams}

\author[DKFZ, HIRO, HDPHYS]{Jennifer~Josephine~Hardt}
\ead{jennifer.hardt@dkfz-heidelberg.de}

\author[DKFZB,KIT]{Alexander~A.~Pryanichnikov}

\author[DKFZ,HIRO,HIT]{Oliver~Jäkel} 

\author[DKFZB,HDPHYS]{Joao~Seco}

\author[DKFZ,HIRO]{Niklas Wahl}

\address[DKFZ]{Division of Medical Physics in Radiation Oncology, German Cancer Research Center -- DKFZ, Im Neuenheimer Feld 280, 69120 Heidelberg, Germany}
\address[HIRO]{Heidelberg Institute for Radiation Oncology  and  National  Center  for  Radiation   Research  in  Oncology, Im Neuenheimer Feld 280, 69120 Heidelberg, Germany}
\address[HDPHYS]{Faculty of Physics and Astronomy, Heidelberg University, Heidelberg, Germany}
\address[DKFZB]{Division of Biomedical Physics in Radiation Oncology German Cancer Research Center -- DKFZ, Im Neuenheimer Feld 280, 69120 Heidelberg, Germany}
\address[HIT]{Heidelberg Ion-Beam Therapy Centrer, Department of Radiation Oncology Heidelberg University Hospital,Germany}
\address[KIT]{Institute of Biomedical Engineering (IBT), Karlsruhe Institute of Technology (KIT), Karlsruhe, Germany}

\begin{document}
\maketitle

\begin{abstract}
\textbf{Background:}
Recently, mixed carbon-helium beams were proposed for range verification in carbon ion therapy: Helium, with three times the range of carbon, serves as an on-line range probe, and is mixed into a therapeutic carbon beam.
\textbf{Purpose:}
Treatment monitoring is of special interest for lung cancer therapy,  however the helium range might not always be sufficient to exit the patient distally. Therefore mixed beam use cases of several patient sites are considered.
\newline
\textbf{Methods:}
An extension to the open-source planning toolkit, \texttt{matRad}, allows for calculation and optimization of mixed beam treatment plans. The use of the mixed beam method in 15 patients with lung cancer, as well as in a prostate and liver case, for various potential beam configurations was investigated. Planning strategies to optimize the residual helium range considering the sensitive energy range of the imaging detector were developed. A strategy involves adding helium to energies whose range is sufficient. Another one is to use range shifters to increase the helium energy and thus range.
\newline
\textbf{Results:}
In most patient cases, the residual helium range of at least one spot is too low. All investigated planning strategies can be used to ensure a high enough helium range while still keeping a low helium dose and a satisfactory total mixed carbon-helium beam dose. The use of range shifters allows for the detection of more spots.
\newline
\textbf{Conclusion:}
The mixed beam method shows promising results for online motioning. The use of range shifters ensures a high enough helium range and more detectable spots, allowing for a wider-spread application.

\end{abstract}


\newpage

\section{Introduction}
Mixed beams are a recent proposition for range-guided particle therapy \citep{mazzucconi_mixed_2018,graeff_oa027_2018}. Hereby two ion species with the same mass-to-charge ratio, for example, fully ionized carbon (\textsuperscript{12}C\textsuperscript{6+}) and helium (\textsuperscript{4}He\textsuperscript{2+}) ions, are accelerated to the same energy per nucleon in a synchrotron-based accelerator. The experimental feasibility of creating such mixed beams was recently demonstrated: \citet{graeff_o072_2024,galonska_first_2024} extract a mixed beam from a single ion source using methane with helium as support gas. \citet{kausel_double_2025,renner_towards_2024} use a double multi-turn injection, injecting first helium and then afterwords carbon into the synchrotron ring.

\begin{figure}[htb]
\centering
\includegraphics{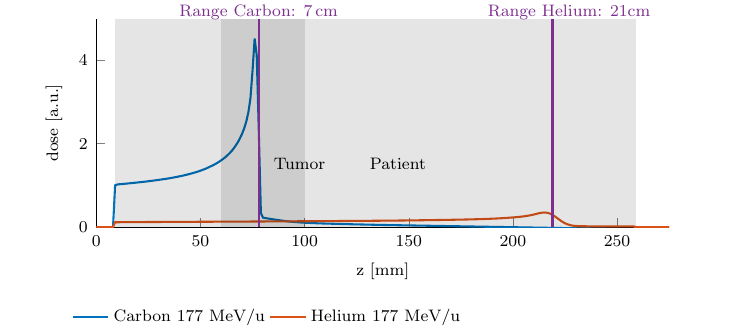}
\caption{Schematic set up of a mixed carbon-helium irradiation, in this example the helium energy is not high enough for the helium ions to exit the patient distally. }
\label{fig:CandHeNotSufficientRange}
\end{figure}

With carbon and helium ions accelerated to the same energy per nucleon, the range of helium ions is about three times the range of carbon ions, as shown in figure \ref{fig:CandHeNotSufficientRange}. Thus, while carbon ions are the primary treatment modality, helium ions, due to their greater range, could exit the patient distally. The residual range of the helium ions could then be measured, offering the potential for an online treatment monitoring to verify the water-equivalent thickness (WET) at the irradiated spot positions. \citet{volz_experimental_2020,mazzucconi_mixed_2018} discussed an additional helium fluence of $10\%$ to ensure a low additional dose to the patient while still achieving a detectable helium signal above the carbon fragment background.

\citet{hardt_potential_2024} demonstrated potential use cases of the mixed carbon-helium beam method in treatment verification, investigating patient positioning verification or breath hold verification. We extended the open source treatment planning tool \texttt{matRad} \citep{wieser_development_2017,abbani_matrad_2024} for mixed beam treatment planning. Including fast pencil beam dose calculation with custom high-energy helium pencil-beam kernels and exportation of the treatment plan to the Monte Carlo system, TOPAS, \citep{perl_topas_2012,faddegon_topas_2020} to allow for simulation of helium radiographs.

The pipeline for simulation of helium radiographs presented by \citet{hardt_potential_2024} modeled the  proton radiography detector by ProtonVDA \citep{dejongh_technical_2021}. This system was recently used for helium radiographs  \citep{pryanichnikov_experimental_2025}, as well as a feasability study for intrafractional motion management \citep{pryanichnikov_feasibility_2025}. This system utilizes two trackers, one proximal and one distal of the patient and an energy detector distal of the patient. However with the high carbon intensities used for radiation therapy, the  proximal or front tracker is expected to be overwhelmed by the used intensity. Therefore the pipeline was extended to  allow for a simulation without front tracker.  The position of the ions are only measured distall of the patient. During image reconstruction the position of the ions on the front tracker is estimated from the beam monitoring system \citep{krah_comprehensive_2018,ordonez_fast_2019}. During our simulations the geometry and materials of the tracking detector is simulated, and the phase space of the primary helium ions at the position of the trackers is saved from which the images are reconstructed.

Taking a closer look at the helium range, Figure \ref{fig:CandHeNotSufficientRange} shows it to be about three times the carbon range at the same energy per nucleon. However, depending on the tumor site the helium range may not be sufficient for the ions to exit the patient distally as also seen in figure \ref{fig:CandHeNotSufficientRange}. In lung cases, common irradiation geometries often require only a small carbon range to reach the tumor resulting in a respectively too short helium range. Complementary, for more deep seated tumors, the high residual helium range of the exiting ions can challenge the sensitive window of the detector. In this work, we investigate the expected residual helium range for several cancer sites, namely prostate, liver and lung. Next to this several helium range strategies are proposed to choose and modify the residual helium range: \textit{EW~He} (energy wise helium),  \textit{const~RaShi} (constant range shifter) and \textit{EW~RaShi} (energy wise range shifter). These strategies will be presented in more detail the following section and should ensure sufficient helium range, to exit the patient distal  but also consider the optimal sensitive range of the used detection system. In short, we propose irradiating only the portion of the energies in the treatment plan with sufficient helium range with a mixed carbon helium beam (\textit{EW~He}). Additionally, we propose adding proximal range shifters to increase the used helium energy, ensuring sufficient range and distal range shifters to reduce the helium range when necessary, ensuring that the residual helium energy is not too high for the detector to measure (\textit{const~RaShi},\textit{EW~RaShi}).

During treatment planning, a raytracing algorithm \citep{siddon_fast_1985} calculates the (water equiavlent path length) WEPL traversed by the particle beam. This allows us to estimate the residual helium range buy subtracting the traversed WEPL from the initial range. 

\section{Materials and Methods}
\subsection{Strategies to optimize the residual Helium range }

\subsubsection{Selection of energies for mixed beam irradiation}

The first investigated strategy (\emph{EW~He}) to avoid mixed-in helium Bragg peaks within the patient is to limit the usage of a mixed carbon-helium beam to energies with sufficiently high helium range. This strategy would first irradiate all carbon-only energies, then switch to a mixed carbon-helium beam and irradiate the remaining energies. To implement this, the minimum residual helium range is calculated for each energy; if it is smaller than a safety margin of \SI{10}{\milli\meter}, the energy is marked for irradiation with a conventional carbon beam only.

\subsubsection{Selection of proximal and distal range shifters}

Since the helium range is three times the carbon range , the range of the primary carbon spot needs to be at least  $1/3$  of the minimum helium range required for the helium ions to distally exit the patient. To ensure that the all primary carbon spots in the target have a large enough water-equivalent range,  a proximal range shifter   ($x_{P}$)  can be used to adjust the proximal material budget.   

Since not every helium energy can be measured by the used detection system, the use of a distal range shifters ($x_{D}$), placed between the patient and the detector is  considered. This range shifter lowers, if needed, the residual helium energy at the detection system for optimal detection properties.



We developed an algorithm to automatically choose the WET of the proximal and distal range shifter based on the residual helium range, and the sensitive range of the chosen detector. Several discrete option for the WET of the proximal ($O_P = \{\SIlist[list-units = single]{0; 5;  15; 25; 35; 45}{\milli\meter}\}$), and for the distal range shifter ($O_D = \{\SIlist[list-units = single]{0; 10; 20; 30; 40; 50; 60; 70; 80; 90; 100; 110; 120; 130; 140; 150}{\milli\meter}\}$) were considered.  The proximal range shifter thicknesses are chosen to be consistent with  range shifters already used in clinics, \citet{wang_investigating_2023}  uses range shifters with a thicknesses of up to \SI{41.2}{\milli\meter}.

First we calculate which proximal range shifters ($x_{P}$) and energy combinations ($E$) are possible, i.e., if the carbon range ($R_C(E)$) is within the target and if the helium range ($R_{He}(E)$) is high enough for helium to traverse the entire patient. We calculate for all energy and proximal range shifter combinations the following, and only consider combinations for which this holds true

\begin{equation}
\mathcal{A}(x^P, E) = 
\begin{cases}
1, & \text{if } \text{WEPL}^{\text{TI}} \leq R_C(E) - x^P \leq \text{WEPL}^{\text{TO}} \\
   & \quad \text{and } R_{He}(E) - \text{WEPL}^{\text{PO}} - x^P \geq R^{\text{min}} \\
0, & \text{otherwise}
\end{cases}
\end{equation}

Hereby $\text{WEPL}^{\text{TI}}$ and $\text{WEPL}^{\text{TO}}$ are the traversed WEPL at the tumor entrance and exit and $\text{WEPL}^{\text{PO}}$ the total WEPL of the patient.  The safety margin of the minimum allowed residual helium range ($R^{\text{min}}$) distal of the patient is \SI{10}{\milli\meter}.

For each of the available combinations we then estimate if the helium ions will be detectable by our chosen detector, i.e if the residual helium range is in the sensitive range of the detector denoted as $[ R^{minD}, R^{maxD}]$. For this we calculate: 

\begin{equation}
\mathcal{D}(x^P, x^D, E) = 
\begin{cases}
1, & \text{if } R^{\text{minD}} \leq R_{He}(E) - \text{WEPL}^{\text{D}} - x^P - x^D \leq R^{\text{maxD}} \\
0, & \text{otherwise}
\end{cases}
\end{equation}

Whereby $\text{WEPL}^{\text{D}}$ is the traversed WEPL at the entrance of the detector. Contrary to $\text{WEPL}^{\text{PO}}$ this could include the WEPL of patient couch.

During investigation the minimal detectable range was set to $R^{\text{minD}} = \SI{7.5}{\milli\meter}$, for the maximum detectable range two imaging detectors, were investigated one with a smaller ($R^{\text{maxD}} = \SI{110}{\milli\meter}$), and one with a larger sensitive range ($R^{\text{maxD}} = \SI{160}{\milli\meter}$).

Two approaches were considered for selecting the optimal range shifter. In the first method, each treatment field has a single fixed proximal and distal range shifter for all delivered energies (\textit{const~RaShi}). In contrast, the second method allows energy-dependent selection of the proximal range shifter (\textit{EW~RaShi}). The concept behind this approach is that the proximal range shifter could be attached to the nozzle, enabling a quick selection of a different thicknesses for a different energy. However, the distal range shifter thickness remains unchanged throughout the delivery of the treatment field.

\paragraph{\textit{Const~RaShi}}

To determine the optimal combination of proximal and distal range shifters for a field, we strive for a high detection percentage while minimizing the WET thickness of the selected range shifters. A greater thickness leads to increased beam broadening and image noise. Therefore, we maximize the weighted sum of of the number of detectable spots and the total thickness of the applied range shifters:

\begin{equation}
	[x^{P\star}, x^{D\star}] = \argmax_{\substack{x^{P} \in O^{P} \\ x^{D} \in O^{D}}}
	\left( \frac{1}{n_1}  \left( \sum\limits_{s\in \mathcal{F}} \mathcal{D}(x^{P}, x^{D}, E_s) \right)
	- \frac{w}{n_2} \left( x^{P} + x^{D} \right)  \right).
	\label{eq:constRaShi}
\end{equation}
The sum is performed for all spots $s$ belonging to the treatment field $\mathcal{F}$. The parameter $w$ serves as a relative weighting factor, of the number of detectable spots to the thickness of the range shifters. The factors $n_1$ and $n_2$ normalize both parts of the function to have equal magnitude

\begin{equation}
	n_1 = \max_{\substack{x^{P} \in O^{P} \\ x^{D} \in O^{D}}}\left(\sum\limits_{s\in \mathcal{F}} \mathcal{D}(x^{P}, x^{D}, E_s)\right)
\end{equation}
\begin{equation}
	n_2 = \max_{\substack{x^{P} \in O^{P} \\ x^{D} \in O^{D}}}\left(x^P + x^D\right)
\end{equation}

\paragraph{\textit{EW~RaShi}}

Selecting the optimal proximal range shifter, while allowing for varying thicknesses across different energies, is more complex, since a given depth in the patient can be reached through multiple energy and proximal range shifter combinations. Therefore we group the spots into intervals ($I_i$) of approximately the same carbon Bragg peak position in the patient($R_C^P = R_C - x^P$). The width of each interval is the chosen longitudinal spot spacing $l$. With this interval $I_i$ is given as
\begin{equation}
    I_i = [\min(R_C^P) + i l, \min(R_C^P) + (i+1) l]  \quad i = 0, 1, \dots, N.
\end{equation}

Whereby the last interval $I_N$ includes the most distal carbon range position ($\max(R_C^P)$).

For each distal range shifter ($x^D$) and carbon range interval ($I_i$), the optimal proximal range shifter ($x^{P\star}_{iD}$) with corresponding energy ($E^{\star}_{iD}$) combination is selected, as the one yielding the highest number of detectable spots.

\begin{equation}
	[x^{P\star}_{iD}, E^{\star}_{iD}] = \argmax_{x^{P} \in O^{P}}
	\left( \sum\limits_{s \in I_i}\mathcal{D}(x^{P}_s, x^{D}, E_s) \right).
\end{equation}

The number of detectable spots for this range interval ($I_i$) with optimized settings, is $\mathcal{D}_{iD}$. The optimal proximal range shifters with their corresponding beam energy is chosen for each distal range shifter option. To determine which distal range shifter ($x^{D\star}$) with corresponding proximal range shifters should be used, the total number of detectable spots is maximized while minimizing the overall thickness of the selected range shifters, similar to the \textit{const~RaShi} slection.

\begin{equation}
	x^{D\star} = \argmax_{x^{D} \in O^{D}} \left( \frac{1}{n_1}  \left(\sum\limits_{i=0}^N \mathcal{D}_{iD} \right) - \frac{w}{n_2}\left( \sum\limits_{i=0}^N  x^{P\star}_{iD} + x^D \right)\right)
\end{equation}

Whereby the normalization is given as

\begin{equation}
	n_1 = \max_{x^{D} \in O^{D}}\left(\sum\limits_{i=0}^N \mathcal{D}_{iD} \right)
\end{equation}
\begin{equation}
	n_2 = \max_{x^{D} \in O^{D}} \left( \sum\limits_{i=0}^N  x^{P\star}_{iD} + x^D \right) 
\end{equation}

\subsubsection{Incorporating proximal range shifters in dose calculation}

To model the additional beam broadening caused by range shifters, Monte Carlo simulations were conducted using TOPAS version 3.9. The simulations employed the following physics list: G4DecayPhysics, G4StoppingPhysics, G4EmExtraPhysics, G4EMStandardPhysics\_option4, G4HadronElasticPhysics, g4h-phy\_QGSP\_BIC\_HP and G4QMDReaction physics. To accurately model helium ions G4BinaryLightIonReaction was activated with the Tripathi cross section data \citep{tripathi_accurate_1999} as modified by \citet{horst_measurement_2019}.

The simulations included carbon and helium beams at five different energies within the carbon treatment energy range ($\SIlist[list-units = single,per-mode=symbol]{88.83;196.23;272.77;339.80;427.44}{\mega\electronvolt\per\atomicmassunit}$). Energy deposition was scored in a cylindrical volume, with a range shifter placed in front of it. Seven different water equivalent thicknesses of the range shifer were used  ($x^P$ = $\SIlist[list-units = single]{5;10;15;20;25;30;35;40;45}{\milli\meter}$). The range shifter was positioned at the nozzle and thus separated from the cylinder's surface at the isocenter by approximately $\SI{1}{\meter}$ of air.

The lateral entrance dose was fitted with a Gaussian function, and the initial beam width was subtracted quadratically to estimate the beam broadening induced by the range shifter  ($\sigma_{RaShi}$). A polynomial function was then fitted to interpolate beam broadening at intermediate energies. During dose calculation, the estimated beam broadening from the range shifter is added to the initial beam width \citep{hong_pencil_1996,wieser_development_2017}.

\subsection{Investigated patient cases}

\begin{figure}[h]
	\centering
	\includegraphics{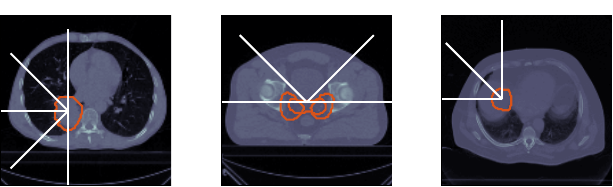}
	\caption{Left: Example lung patient from the lung data set, outlined is the target volume (\protect\tikz[baseline=-0.5ex]\protect\draw[very thick, myorange] (0,0) -- (0.5,0);) and the gantry angles for which the residual helium range was investigated. Middle: Prostate case, Right: Liver case}
	\label{fig:allPatientsOverview}
\end{figure}

We analyzed the residual helium range in $15$ lung cancer patient cases \citep{hugo_data_2016}, considering five different gantry angles: \SIlist{0; 180}{\degree}, as well as \SIlist{45; 90; 135}{\degree} for tumors in the right lung or \SIlist{315; 270; 225}{\degree} for tumors in the left lung. An example patient is shown in figure \ref{fig:allPatientsOverview}.

Additionally, we conducted a detailed analysis of one patient (No.~114) by generating treatment plans for a gantry angle of \SI{270} across all residual helium range strategies. The dose influence matrix was calculated using a regular spot grid with lateral spacing of \SI{5}{\milli\meter}  and a longitudinal  spacing of approximately \SI{2}{\milli\meter} (depending on available energies) on a \SI{3}{\milli\meter^3} dose grid. The RBE-weighted carbon ion dose (LEM I) was optimized to  \SI{2.3}{\gray}  per fraction for the target PTV.

Furthermore, we examined the residual range in a prostate cancer patient \citep{craft_shared_2014}, with both a low-dose PTV (\SI{56}{\gray}) and a high-dose PTV (\SI{68}{\gray}), evaluating gantry angles of \SIlist{45; 90; 270; 315}{\degree}. For a liver cancer patient \citep{craft_shared_2014}, we investigated gantry angles of \SIlist{0; 315; 270}{\degree}. 

For the prostate and liver cases, we assigned a relative weighting factor $w$ of $0.25$ (equation \ref{eq:constRaShi}), prioritizing more the number of detectable spots. For the lung cases, this weighting factor was set to $0.5$. An overview of the investigated tumor sites and corresponding gantry angles is provided in figure \ref{fig:allPatientsOverview}.

\section{Results}

\subsection{Residual Range analysis for different cancer sites}

\subsubsection{Lung cases}

\begin{figure*}
	\centering
	\hspace{-1.5cm}
	\begin{subfigure}[b]{0.40\textwidth}
		\centering
		\includegraphics{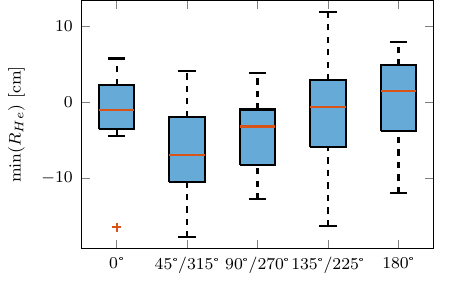}
	\end{subfigure}
	\hspace{1cm}
	\begin{subfigure}[b]{0.40\textwidth}
		\centering
		\includegraphics{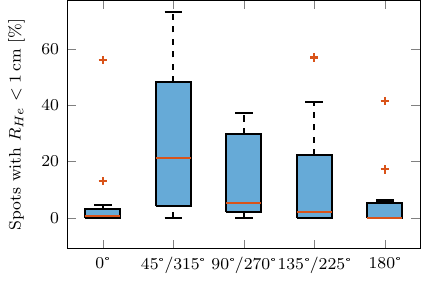}
	\end{subfigure}
	\caption{Left: Box plot summarizing the minimum residual helium range of each lung patient for the different gantry angles. Right: Box plot summarizing the percentage of spots in each treatment plan with a residual helium range smaller than \SI{1}{\centi\meter} for each gantry angle.}
	\label{fig:minHeRangeLung}
\end{figure*}

During the investigation of the residual helium range, none of the previously mentioned helium range strategies were applied; instead, conventional treatment plans were created, as the first phase of this study aimed to evaluate the necessity of such strategies. Figure \ref{fig:minHeRangeLung} provides an overview of the minimum residual helium range for each investigated gantry angle and patient case, along with the percentage of spots where the residual helium range is less than \SI{1}{\centi\meter}. This serves as an indication of the severity, helping to determine whether only a small or large fraction of spots are identified as having insufficient range. Negative values indicate spots where the helium ions lacked sufficient energy to exit the patient distally.

Among the examined angles, $\SI{0}{\degree}$ and $\SI{180}{\degree}$ appear to be the most suitable, for mixed beam irradiation without range shifters, as they offer a higher minimum residual helium range and a lower percentage of spots with insufficient range. However, $\SI{0}{\degree}$ may be preferable to $\SI{180}{\degree}$ since it avoids irradiation through the patient couch. Despite this, for $9$ out of the $15$ patients analyzed, the minimum residual helium range for a gantry angle of $\SI{0}{\degree}$ was still below the \SI{1}{\centi\meter} safety margin. Furthermore, in $6$ of the $15$ cases, none of the investigated angles provided sufficient range. The most frequently available angle was \SI{180}{\degree} ($8$/$15$) followed by \SI{0}{\degree}  and $135/225$\si{\degree} ($6$/$15$). 

\begin{figure}[h]
	\centering
	\includegraphics{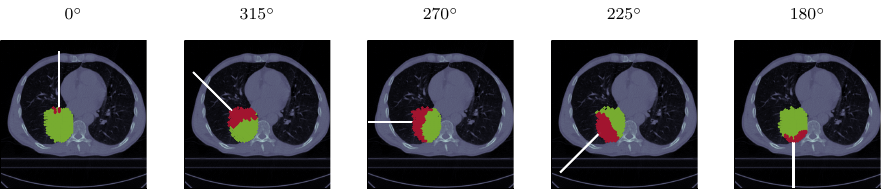}
	\caption{Axial CT slice for patient No.114 for different gantry angles. The overlay highlights the spots with sufficiently large residual helium range in green (\protect\tikz[baseline=-0.5ex]\protect\draw[fill=mygreen] (0,0) circle(0.1cm);) and the ones with insufficient residual helium range in read (\protect\tikz[baseline=-0.5ex]\protect\draw[fill=myred] (0,0) circle(0.1cm);).}
	\label{fig:LungAllAnglesSpotMap}
\end{figure}

Figure \ref{fig:LungAllAnglesSpotMap} uses an exemplary lung patient (No.~114) to illustrate  the distribution of spots with sufficient and not sufficient helium range. For the $\SI{315}{\degree}$  gantry angle, it is particularly evident that the helium beam is most affected in regions where it must pass through the spine.

\subsubsection{Prostate Case}  

In the prostate case, the residual helium range was examined for four gantry angles. Compared to the lung case, much fewer spots were affected by insufficient residuel helium range. Specifically, only \SI{1.1}{\percent} of spots exhibited insufficient helium range for the \SI{90}{\degree} angle, which represented the highest percentage of spots with insufficient helium range among the four investigated angles.   For the other angles, the percentages were \SI{0.9}{\percent} ( \SI{270}{\degree}), \SI{0.2}{\percent} ( \SI{45}{\degree}), \SI{0}{\percent} ( \SI{315}{\degree}). Despite the prostate being centrally located within the patient, which might suggest that residual helium range would not be a concern, it still is.

\subsubsection{Liver Case}

For the liver case, the residual helium range was analyzed for three gantry angles. All three angles exhibited spots with insufficient residual helium range. For the  \SI{0}{\degree} gantry angle, only \SI{1}{\percent} of the spots were flagged as insufficient, whereas \SI{47}{\percent} and \SI{38}{\percent} of spots were flagged as insufficient in the  \SI{270}{\degree} and  \SI{315}{\degree} angles, respectively.

\subsection{Beam broadening due to range shifters}

\begin{figure}[h]
	\centering
	\includegraphics{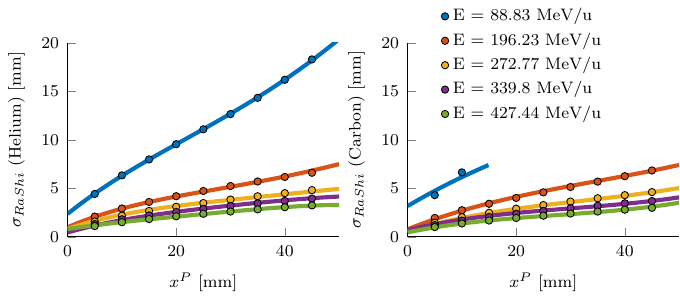}
	\caption{Estimated  (\protect\tikz[baseline=-0.5ex]\protect\draw[fill=black] (0,0) circle(0.1cm);) and fitted  (\protect\tikz[baseline=-0.5ex]\protect\draw[very thick,black] (0,0) -- (0.5,0);) Beam widening due to the use of a proximal range shifter of varying thickens for helium (left) and carbon (right). }
	\label{fig:HeAndCarbonRashiSigma}
\end{figure}

Figure \ref{fig:HeAndCarbonRashiSigma} shows the estimated beam widening $\sigma_{RaShi}$ caused by the the use of proximal range shifter as well as the interpolated function. Especially for low energy's and a thick range shifter the broadening is large. However, this combination is infrequently used in treatment plans.   Normally a thicker range shifter will be used with a higher helium energy. Not for all carbon energy's and proximal range shifter combinations the widening was calculated, for the lowest energy the used range shifter were sometimes thicker than the corresponding carbon range. This would also mean that these energys would not be used in a treatment plan, and thus it is nor necessary to calculate the helium widening in these cases, however this was done for completeness. This beam broadening was used in the following dose calculations.

\subsection{Strategies to optimize the residual helium range}

In general, all residual helium range strategies effectively ensure a sufficient helium range. The strategies are compared based on the percentage of detectable spots. Helium ions may not be detectable for two main reasons: they are lost in the distal range shifter or their residual energy is too high for the detector to measure.

\subsubsection{Lung cases}

\begin{figure}[h]
	\centering
	\includegraphics{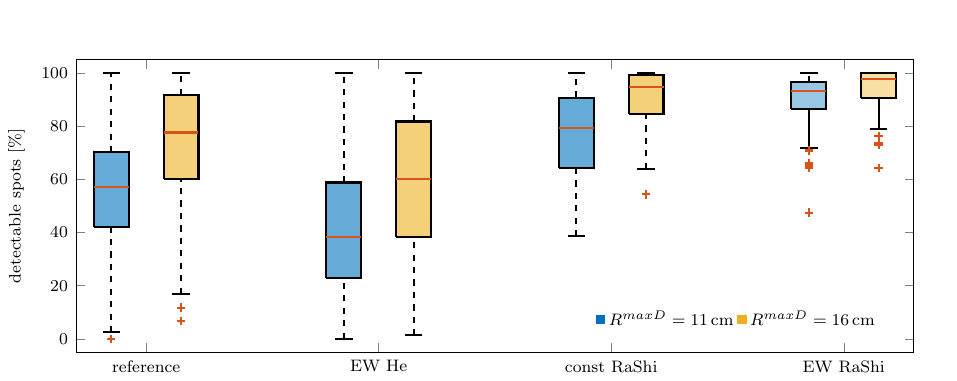}
	\caption{ Box plot summarizing the percentage of detectable spots in the lung treatment plans for a detector with a sensitive range up to \SI{11}{\centi\meter} and a bigger detector with a sensitive range up to \SI{16}{\centi\meter}.}
	\label{fig:isDetectLungBox}
\end{figure}

Figure \ref{fig:isDetectLungBox} presents the percentage of detectable spots across all treatment angles and patients for the presented residual helium range strategies. The calculation was conducted for two imaging detectors, one with a smaller and one with a larger sensitive range. As a reference, a treatment plan was calculated without applying any residual helium range strategy, meaning that this plan may include spots with an insufficient helium range. As expected, the larger detector with wider acceptable residual helium range allows for a larger percentage of detectable spots. Still, the difference between the two detectors is relatively small for the best-performing method, \textit{EW~RaShi} and averages $5$ pp. Among the strategies, \textit{EW~He} performs the worst, even falling below the reference plan. When comparing the two range shifter strategies, the added flexibility of the \textit{EW~RaShi} method results in an average increase in detectable spots of $12$ pp for the detector with the smaller sensitive range and $4$ pp for the larger one. However, for the  detector with the larger sensitive range, smaller distal range shifters are used on average, which would enhance image quality.

\subsubsection{Lung case No.114}

To look at the different strategies in more detail, the delivered dose was optimized and calculated for an example patient. Table \ref{tab:Detectbaility Lung} lists the percentage of detectable helium spots, and since the optimal fluence of each spot was now calculated, the percentage of detectable helium ions was also calculated. When calculating the percentage of detectable helium ions for the  \textit{EW~He} method, not the total amount of delivered helium ions was used, but \SI{10}{\percent} which is the carbon-helium ratio, of the carbon ions used. This allows a better comparison between all strategies. For the reference and  \textit{EW~He} strategies, the percentage of detectable helium ions increases while the percentage of detectable ions decreases for the  \textit{const~RaShi} strategy, when compared to the percentage of detectable spots.  Due to higher fluence of the  intermediate energies, which are detetectable by the reference plan, the reference and \textit{const~RaShi} have the same percentage of detectable helium ions, although the percentage of detectable spots is significantly higher for the \textit{const~RaShi} strategy. A range shifter of only \SI{1.5}{\centi\meter} (\textit{const~RaShi}) already allows the detection of $24$ pp more helium ions than with the \textit{EW~He} method and as many helium ions as with the reference plan, but with sufficient residual helium range.

\begin{table*}
	\centering
	\begin{tabular}{ c | c c c c}
		\toprule
		& reference & \textit{EW~He} & \textit{const~RaShi} & \textit{EW~RaShi}\\
		\midrule
		detectable spots [\%] & 55& 35&  76& 93\\ 
		detectable Helium ions [\%]& 69 & 46& 70 & 93\\  
		\bottomrule   
	\end{tabular}
	\caption{Percentage of detectable spots and helium ions for each method}
	\label{tab:Detectbaility Lung} 
\end{table*}

\begin{figure}[h]
	\centering
	\includegraphics{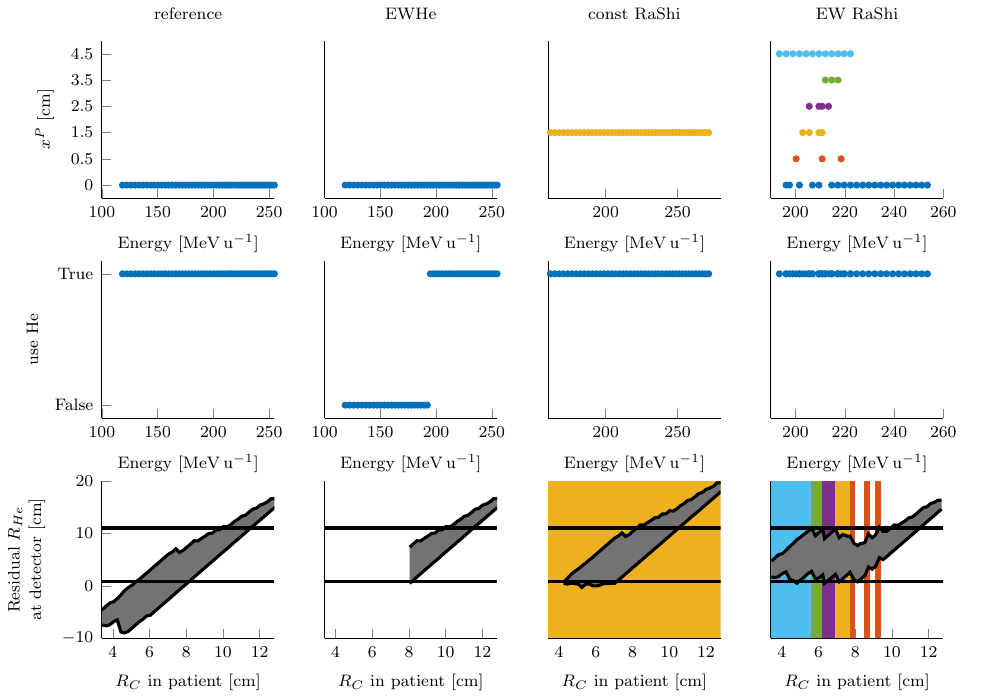}
	\caption{For each strategy (by column) the top row shows the proximal range shifter thickness per energy in the treatment plan. In the middle row illustrates the decision of mixing helium (True) into the carbon beam or not (False) for each energy. The bottom row displays the residual helium range at the detector for each carbon spot position in the patient. Highlighted is the minimum (\SI{0.75}{\centi\meter}) and maximum (\SI{11}{\centi\meter}) detectable range and the used proximal range shifter thickens. Whereby  \textcolor{mylightblue}{\rule{0.5em}{0.5em}} represents a proximal range shifter thickness of  $\SI{45}{\milli\meter}$, \textcolor{mygreen}{\rule{0.5em}{0.5em}} $\SI{35}{\milli\meter}$, \textcolor{mypurple}{\rule{0.5em}{0.5em}} $\SI{25}{\milli\meter}$, \textcolor{myyelllow}{\rule{0.5em}{0.5em}} $\SI{15}{\milli\meter}$ and  \textcolor{myorange}{\rule{0.5em}{0.5em}} $\SI{5}{\milli\meter}$}
	\label{fig:raShiSelectionExamplePatient}
\end{figure}

 Figure \ref{fig:raShiSelectionExamplePatient} provides a closer look at the selected proximal range shifter thickness, which treatment energies used helium, and the last row shows the residual helium range at the detector versus the carbon range in the patient, overlaid with the used proximal range shifter thickness. No distal range shifters were used in this case and the smaller detector ($R^{maxD} = \SI{11}{\centi\meter}$) was used. The \textit{EW~RaShi} strategy uses thicker proximal range shifters than the \textit{const~RaShi} strategy. Looking at the lower right plot, it is clear that the thickest proximal range shifter is used for the lowest carbon range in the patient and the lowest energy, and then the thickness of the proximal range shifter decreases. Note how the proximal range shifter is used to "push back" the residual helium range into the sensitive range of the imaging detector. So while the \textit{EW~RaShi} method has $23$ pp more detectable helium ions than the \textit{const~RaShi}, it comes at the cost of a thicker proximal range shifter. For the \textit{EW~He} strategy, helium is used only with high energys, where the helium range is sufficient. Also noteworthy for the \textit{const~RaShi} method is that the smallest carbon range is not irradiated. These spots must have too low helium range and were therefore excluded, but a thicker range shifter would have come at the cost of a reduced number of detectable spots.

\begin{figure}[p]
	\centering
	\includegraphics{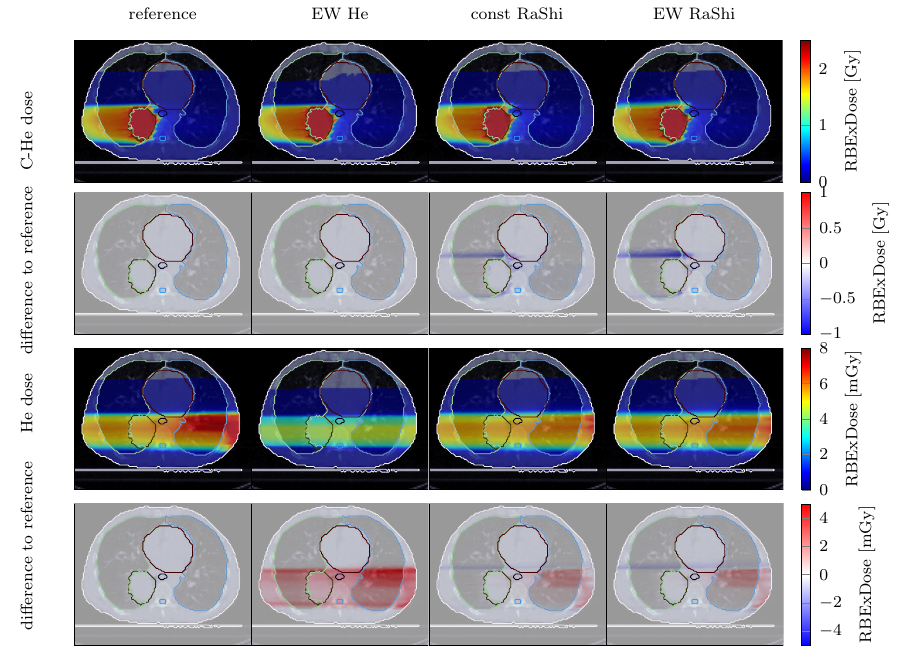}
	\caption{ Axial dose slices for each strategy. Top: total mixed carbon-helium RBE weighted doses and bellow the difference to the reference plan. Bottom: Helium RBE weighted dose and the difference to the reference plan, please note the mGy scale in this case.}
	\label{fig:DoseComparisonLungPatient}
	\centering
	\includegraphics{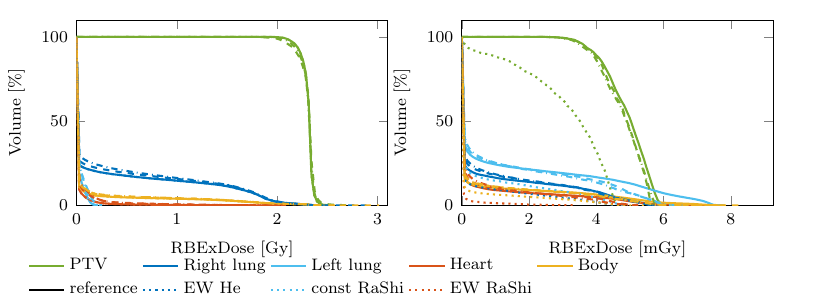}
	\caption{Dose-volume histogram for the different strategy's. The left DVH shows the total mixed carbon-helium RBE weighted dose, while the right DVH singles out the helium dose (on a \si{\milli\gray} scale).}
	\label{fig:DVHComparisonLungPatient}
\end{figure}

The mixed carbon-helium and helium doses for all strategies are shown in figure \ref{fig:DoseComparisonLungPatient} with the respective difference from the reference plan. Figure \ref{fig:DVHComparisonLungPatient} shows the corresponding carbon-helium and helium dose-volume-histograms. The residual helium range strategies reduce the delivered helium dose. This is particularly noticeable for the \textit{EW~He} method with a significant reduction in helium dose to the PTV, left lung, heart and body. However, since the helium dose contributes little to the total dose, the dose reduction is barely visible in the mixed-dose DVH. 

Strategies involving range shifters introduce increased lateral scatter as visible in figure \ref{fig:DoseComparisonLungPatient} as a widened dose profile and resulting increase in dose to the right lung. The mean dose increases from \SI{0.28}{\gray} (reference) to \SI{0.31}{\gray} (\textit{const~RaShi}) and \SI{0.33}{\gray}  (\textit{EW~RaShi}). For these plans, there is also a reduction in target coverage: the $D_{95}$ value of the PTV decreases from \SI{2.19}{\gray} (Reference) to \SI{2.13}{\gray} (\textit{const~RaShi}) and \SI{2.16}{\gray} (\textit{EW~RaShi}). In general, using range shifters increases the total delivered dose of the plans over the reference plan, in this case \SI{10}{\percent} (\textit{const~RaShi}) and  \SI{13}{\percent} (\textit{EW~RaShi}). For all plans, the contribution of helium to the total RBE-weighted dose is less than \SI{1}{\percent}, being \SI{0.57}{\percent} (reference), \SI{0.26}{\percent}(\textit{EW~He}), \SI{0.50}{\percent} (\textit{const~RaShi}) and \SI{0.48}{\percent} (\textit{EW~RaShi}).

\subsubsection{Prostate case}

The prostate case exhibits large residual helium ranges for some energy layers and gantry angles, with the residual range reaching up to \SI{45}{\centi\meter} at a gantry angle of \SI{90}{\degree}.  These large residual helium ranges complicate the detection. The \textit{EW~RaShi} method has the highest percentage of detectable spots. Still for opposing beams (\SI{90}{\degree} ,\SI{270}{\degree} ) it is only \SI{56}{\percent} for the detector with the smaller sensitive range. For the detector with the bigger sensitive range, it increases to \SI{69}{\percent}. The other two investigated angles (\SI{45}{\degree}, \SI{325}{\degree}) have a higher percentage of detectable spots. For the smaller detection system \SI{72}{\percent} can be detected, for the larger system this increases to \SI{89}{\percent}. In the prostate case the residual helium range strategies selected distal range shifters.

\subsubsection{Liver case}
For the liver case, the \textit{EW~RaShi} continued to show the highest percentage of detectable spots; up to \SI{89}{\percent} were detectable for a gantry angle of \SI{270}{\degree} using the detector with reduced range sensitivity. Using the detector with the bigger sensitive range enabled detection of \SI{97}{\percent} of the spots. Only for one gantry angle, \SI{0}{\degree}, a distal range shifter was used.

\subsection{Helium radiographs with range shifter}

\begin{figure*}
	\centering
	\begin{subfigure}[b]{0.33\textwidth}
		\centering
		\includegraphics{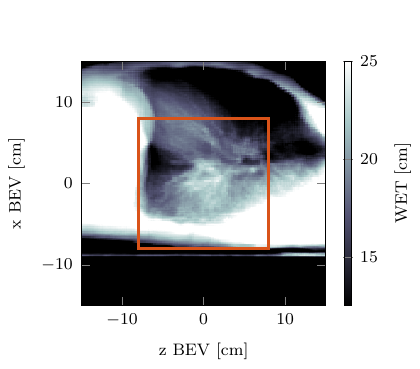}
	\end{subfigure}
	\hspace{2cm}
	\begin{subfigure}[b]{0.33\textwidth}
		\centering
		\includegraphics{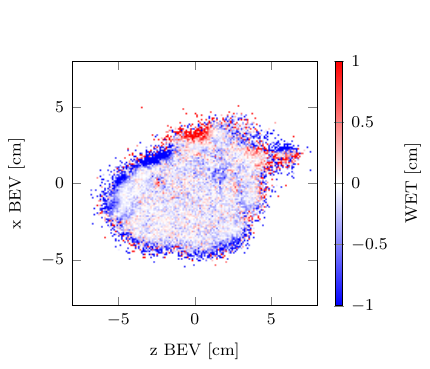}
	\end{subfigure}
	\begin{subfigure}[b]{0.33\textwidth}
		\setlength\fwidth{1\textwidth}
		\centering
		\includegraphics{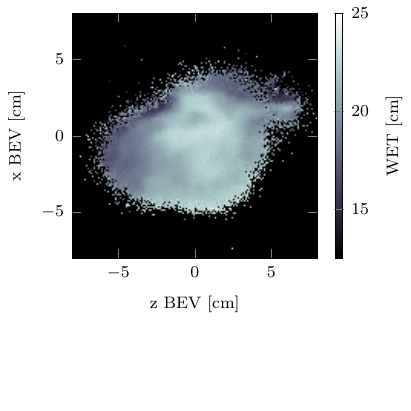}
	\end{subfigure}
	\hspace{2cm}
	\begin{subfigure}[b]{0.33\textwidth}
		\centering
		\includegraphics{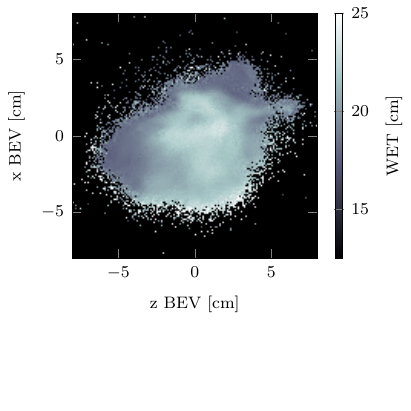}
	\end{subfigure}
	\vspace{-2cm}
	\caption{The top left shows the projected CT (for the \SI{90}{\degree} gantry angle), with the area highlighted (\protect\tikz[baseline=-0.5ex]\protect\draw[very thick, myorange] (0,0) -- (0.5,0);) chosen for the mixed-beam radiographs in the other subfigures. Bottom left: Simulated helium radiograph (\SI{197.58}{\MeVperU}, \SI{0}{\centi\meter}). Bottom right: Simulated helium radiograph (\SI{217.25}{\MeVperU}, \SI{1.5}{\centi\meter}). Top right: Difference Image of both radiographs}
	\label{fig:HeRadComparisonLung}
\end{figure*}

In figure \ref{fig:HeRadComparisonLung}  two helium radiographs are compared, one acquired from the reference plan and the other acquired from a \textit{const~RaShi} plan. The energies of the helium radiographs where chosen so that both plans with and without range shifter, irradiate approximately the same crossecton of the tumor. The difference of both radiographs also indicates an increase in image noise when adding range shifters.

\section{Discussion}

This study used a mixed carbon-helium beam treatment planning framework to investigate the applicability and limits of the mixed beam method imposed by residual helium ranges at different patient sites, with a focus on lung patients. Residual helium range is a limiting factor in the selection of patients and treatment angles, as not all combinations are sensible without mitigating strategies. Two key considerations were identified: first, the helium ions must have sufficient range to exit the patient distally; second, the residual range should be in the sensitive range of the detector, as if the range is too high, the helium ions cannot be detected.

The analysis of the helium range, without any applied helium range strategy, showed insufficient helium range for at least one energy layer in almost all cases, independent of the gantry angle. The sensitive range of the employed detector was found to be a critical factor in ensuring good detection properties. This is particularly relevant for the prostate case, where the helium ions had a greater residual helium range, requiring a larger sensitive range of \SI{16}{\centi\meter} or more. For the lung sites with, in general, a smaller residual helium range, it seems that the smaller detector ($R^{maxD}$ =  \SI{11}{\centi\meter}) is sufficient to enable good detection properties, when a residual helium range strategy such as the \textit{EW~RaShi} is used. Under this approach, the mean increase of detectable spots, between both investigated detectors was only $5$ pp. However, the maximum increase of detectable spots was $26$ pp. Therefore, depending on the patient site and gantry angle, a detector with a larger sensitive range might still be sensible.  An argument for a detector with a larger sensitive range is also that thinner distal range shifters where used with this detector, increasing image quality.

The detectors used by \citet{volz_experimental_2020, mazzucconi_mixed_2018} for the experimental exploration of the mixed beam method covered a total WET of $\sim \SI{127}{\milli\meter}$ and $\sim \SI{180}{\milli\meter}$. While most detectors currently used in ion imaging are calorimeters or range telescopes, measuring residual energy or range, other detection systems are also used. \citet{gehrke_proof_2018} uses a thin silicon pixel detector for helium radiography, which measures deposited energy. However, this system has a very sensitive WET range, as the most accurate WET measurement is achieved in the steep gradient of the Bragg peak. Therefore, for accurate imaging of complex objects, multiple energies are used \citep{metzner_energy_2024}. Another approach measures the time-of-flight (TOF) of an ion between two or more tracking units, from which the velocity and kinetic energy can be calculated. The sensitive WET range of such  a detector depends on the distance between the tracking units and their time resolution \citep{krah_relative_2022,ulrich-pur_first_2024}. A detector designed for mixed beam radiotherapy should be capable of handling high particle fluxes, offer a large sensitive range, and provide particle identification to distinguish helium signals from those of fragments.

For the lung cases, gantry angles of $\SI{0}{\degree}$ and  $\SI{180}{\degree}$ appear most suited for radiation therapy, as only a small portion of the spots exhibit insufficient residual helium range. These angles could also be combined in a two-field treatment plan, where the upper region of the tumor is treated with a $\SI{180}{\degree}$ gantry angle, while the lower region is irradiated with a $\SI{0}{\degree}$ gantry angle, treating only spots with sufficient helium range from each angle. In the prostate case, a similar strategy could be used with opposing gantry angles of  \SI{90}{\degree} and  \SI{270}{\degree}. However, the limitation of this strategy is that only very few centers use carbon gantries like, e.\,g., the HIT (Heidelberg Ion Therapy Center). Most facilities use fixed beam lines instead and are limited to horizontal and vertical orientations, with a few offering an oblique beam line  \footnotemark\footnotetext{\url{https://www.ptcog.site/index.php/facilities-in-operation-public}}.  A fixed beam line may offer an advantage over a gantry for mixed beam irradiation, as it could simplify the setup of a distal range shifter.  For more flexibility with treatment angles in the mixed beam irradiation, upright particle therapy \citep{volz_opportunities_2024} would be beneficial.

A disadvantage of using range shifters to optimize the residual helium range is that they introduce additional scattering, leading to beam broadening and an increased dose delivered to the patient. In addition to increased scattering, they lead to the production of more fragments. Furthermore, as illustrated in figure \ref{fig:HeRadComparisonLung} the use of range shifters increases image noise in the simulated radiographs. Experimental investigation of the detection properties, especially when thick distal range shifters are used is necessary.

\section{Conclusion}

The residual helium range in mixed beam treatment plans was evaluated across several beam angles in lung cases, as well as in a prostate and a liver case, revealing that the helium range was too low in most scenarios. Three different strategies were applied to ensure sufficient helium range. The \textit{EW~He} strategy uses a mixed carbon-helium beam only for energies with sufficient helium range, otherwise a pure carbon beam is used. The \textit{const~RaShi} and \textit{EW~RaShi} incorporate proximal and distal range shifters to ensure sufficient helium range while also optimizing detection capabilities by incorporating the sensitive range of the detector in the selection of the range shifters. 

Overall, all evaluated residual helium range strategies successfully ensure sufficient helium range. Strategies involving range shifters further increase the number of detectable spots. These strategies allow more flexibility in the chosen treatment angle. All of this increases the clinical usability of the mixed carbon-helium beam method, which shows promising potential for range verification, offering online beams-eye-view information, and allowing for the visualization of the treated patient's anatomy.

\section*{Acknowledgments}

This work was funded by the Deutsche Forschungsgemeinschaft
(DFG, German Research Foundation) – Project No. 457509854.

\section*{Conflict of interest}
The authors declare that they have no known competing financial interests or personal relationships that could have appeared to influence the work reported in this paper.

\section{References\label{bibby}}

\printbibliography[heading = none]


\end{document}